\documentclass[sigconf, screen, nonacm]{acmart}

\makeatletter
\def\ACM@jdslogo{}
\def\@ACM@doclicense{}
\makeatother
\AtBeginDocument{%
  }

\setcopyright{none}

\usepackage{hyperref}
\usepackage{fancyhdr}
\usepackage{pifont}
\usepackage{algorithm}
\usepackage{algpseudocode}
\usepackage{multirow}
\usepackage{tabularx}
\newcolumntype{C}{>{\centering\arraybackslash}X}
\usepackage[table]{xcolor}
\usepackage{tikz}
\usetikzlibrary{patterns, positioning, arrows.meta, calc}
\definecolor{blue}{RGB}{33,113,181}   
\definecolor{green}{RGB}{35,139,69}   
\definecolor{red}{RGB}{180,15,32}     
\newcommand{\blue}[1]{\textcolor{blue}{\textsc{#1}}}
\newcommand{\green}[1]{\textcolor{green}{\textsc{#1}}}

\newcommand{\best}[1]{\textcolor{green}{\bf #1}}

\makeatletter
\def\@ACM@doclicense{}
\makeatother

\begin{document}

\title{NPU Design for Diffusion Language Model Inference}
\renewcommand{\shortauthors}{Lou et al.}
\hypersetup{
  pdftitle={NPU Design for Diffusion Language Model Inference},
  pdfauthor={Binglei Lou, Haoran Wu, Kevin Lau, Gregor MacDonald, Jiayi Nie, Yao Lai, Can Xiao, Xuan Guo, Jianyi Cheng, Rika Antonova, Robert Mullins, Aaron Zhao}
}

\renewcommand{\authorsaddresses}{}
\fancyhead[RE]{}

\author[Lou et al.]{%
\normalsize
Binglei Lou\textsuperscript{$*$} \quad
Haoran Wu\textsuperscript{$\dagger$} \quad
Kevin Lau\textsuperscript{$*$} \quad
Gregor MacDonald\textsuperscript{$*$} \quad
Jiayi Nie\textsuperscript{$\dagger$} \quad
Yao Lai\textsuperscript{$\dagger$} \quad
Can Xiao\textsuperscript{$*$} \quad 
Xuan Guo\textsuperscript{$*$} \quad \\
Jianyi Cheng\textsuperscript{$\S$} \quad
Rika Antonova\textsuperscript{$\dagger$} \quad
Robert Mullins\textsuperscript{$\dagger$} \quad
Aaron Zhao\textsuperscript{$*$} \quad
}
\affiliation{%
\institution{%
\normalsize
\textsuperscript{$*$}Imperial College London,
\textsuperscript{$\S$}University of Edinburgh,
\textsuperscript{$\dagger$}University of Cambridge
}
\country{}}

\begin{abstract}
Diffusion-based LLMs (dLLMs) fundamentally depart from traditional autoregressive (AR) LLM inference: they leverage bidirectional attention, block-wise KV cache refreshing, cross-step reuse, and a non-GEMM-centric sampling phase. These characteristics make current dLLMs incompatible with most existing NPUs, as their inference patterns, in particular the reduction-heavy, top-$k$-driven sampling stage, demand new ISA and memory hierarchy support beyond that of AR accelerators. 
In addition, the blocked diffusion KV cache breaks from the append-only paradigm assumed by AR NPUs, and conventional AR-derived KV quantization schemes were designed for static activation distributions and do not account for the step-wise distribution shifts introduced by iterative block-wise refinement in dLLMs.

In this paper, we introduce the first NPU accelerator specifically designed for dLLMs. It delivers: a dLLM-oriented ISA and compiler; a hardware-optimized execution model for both the transformer inference and diffusion sampling used in dLLMs; a novel Block-Adaptive Online Smoothing (BAOS) for quantizing KV cache in dLLMs; and a complete RTL implementation synthesized in 7nm. To evaluate and validate our design, we introduce a tri-path simulation framework that comprises analytical, cycle-accurate, and accuracy simulators, together with cross-validations against physical hardware. 
The full NPU stack, including ISA, simulation tools, and quantization software, will be open-sourced upon acceptance.

\end{abstract}


\maketitle

\section{Introduction}
\label{se:Introduction}

A decade of hardware innovation for autoregressive (AR) large language models has produced a broad ecosystem of AI accelerators, optimized across both the compute-bound prefill phase and the memory-bandwidth-bound decode phase (e.g., the NVIDIA Vera Rubin platform~\cite{rubin}). Yet diffusion-based LLMs (dLLMs)~\cite{llada,ye2025dream,lou2023discrete} have recently demonstrated competitive generation quality with a fundamentally different computational structure, and no dedicated hardware accelerator for dLLM inference exists. AR accelerators~\cite{microscopiq,olive,figna,plena}, whether optimized for prefill, decode, or disaggregated across both, share architectural assumptions that dLLMs collectively violate: append-only KV caching, a single-pass transformer with no iterative denoising, and GEMM-centric compute throughout. In dLLMs, every diffusion step is a full-sequence prefill-like pass; the KV cache undergoes block-wise refresh and cross-step reuse rather than monotonic growth; and the sampling stage is reduction-heavy with no AR analogue. Adapting existing AR accelerators is therefore not a straightforward extension: the memory hierarchy, instruction set, KV cache management, and quantization pipeline must all be reconsidered.

\begin{figure}[t]
    \centering
    \includegraphics[width=1.0\linewidth]{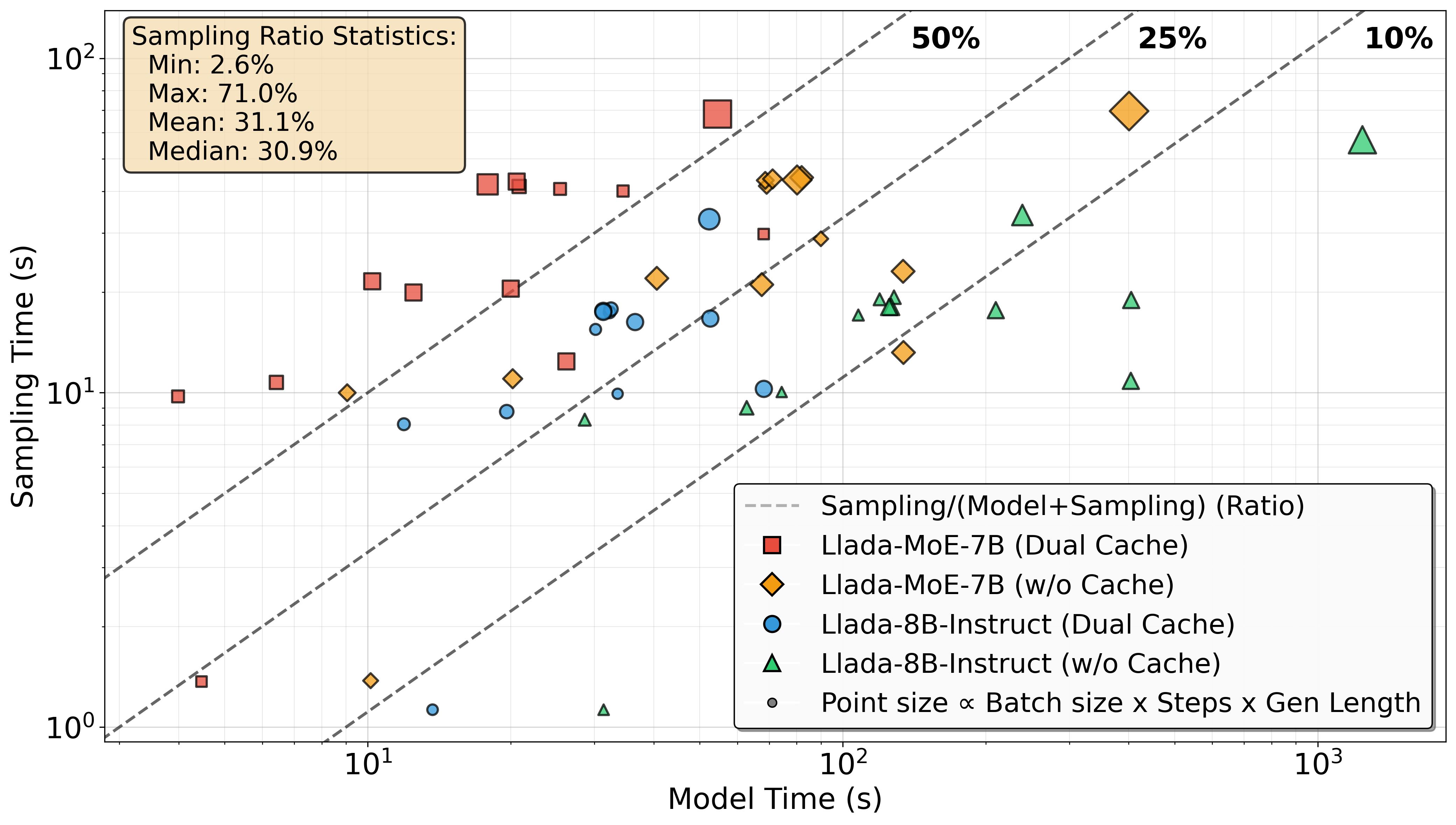}
    \caption{Latency breakdown (model and sampling) of LLaDA-8B-Instruct and LLaDA-MoE-7B-A1B on an NVIDIA A6000 GPU under the reference software configuration, profiled across batch sizes 1--32, denoising steps 1--32, generation lengths 64--1024, and block sizes 8--64 (dInfer framework~\cite{dinfer} with optimized vLLM backends). The sampling stage reaches up to 71\% of end-to-end latency under MoE and dual KV-cache configurations.}
    \label{fg:intro}
\end{figure}

Among these mismatches, the sampling stage represents a non-trivial and previously unaddressed hardware bottleneck. Profiling representative dLLM workloads on GPU under the reference software configuration (FP64 sampling precision~\cite{llada}) shows that sampling accounts for up to 71\% of end-to-end inference latency (Fig.~\ref{fg:intro}). NPU deployment makes reduced-precision sampling a necessity rather than an option. We benchmark FP64$\to$BF16$\to$MXFP8 across LLaDA-series models~\cite{llada,wu2025fast} (Section~\ref{subsec:kv_quant}): MXFP8 preserves generation quality, and the sampling percentage drops to under 10\% of end-to-end latency. Beyond precision, the computational structure of sampling has no AR analogue. Unmasking each token requires a diffusion denoising that demands fused max-with-index, streaming top-$k$, and integer masked token updates that do not exist in standard ISAs. The three data types produced at each step (logit vectors, FP confidence scalars, integer token indices) further require physical isolation in on-chip memory domains to avoid address-decoder contention.


KV cache quantization is another challenge. AR-derived methods such as Hadamard-based rotation~\cite{quarot} and per-channel smoothing~\cite{p3llm} were designed under the assumption that KV activation distributions remain stable across decode steps, which does not hold for dLLMs: multi-step block-wise refinement progressively unmasks tokens and shifts channel-wise outlier statistics across denoising steps. At the same time, Fast-dLLM-style block decoding~\cite{wu2025fast,wu2025fast-v2} introduces a natural calibration opportunity through its per-block warm-step recomputation. No prior method exploits this dLLM-specific structure, leaving dLLM KV quantization without an effective hardware-deployable solution.

\begin{figure}[t]
    \centering
    \includegraphics[width=1.0\linewidth]{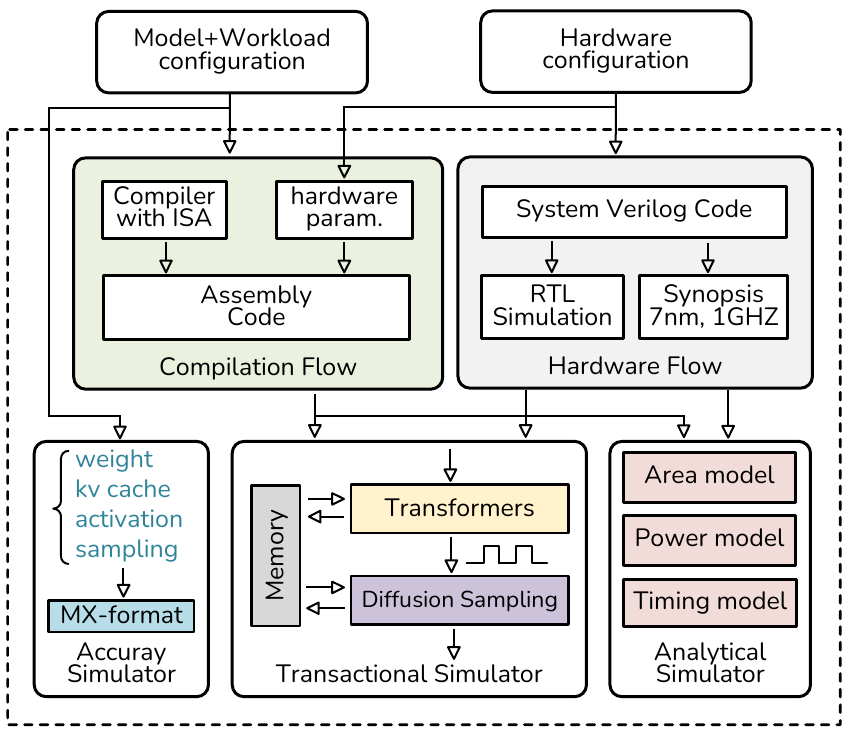}
    \caption{Overview of the Dart NPU for dLLM Inference.}
    \label{fg:framework_overview}
\end{figure}

We present \textbf{DART}, the first configurable NPU hardware platform for dLLM inference (Fig.~\ref{fg:framework_overview}). DART provides a parameterizable design covering the full dLLM execution stack: the transformer forward pass, diffusion sampling, and block-wise KV caching, targeting both dense and MoE dLLM variants. A tri-path simulation framework comprising analytical, cycle-accurate, and accuracy simulators, cross-validated against AMD HBM2e physical measurements and Verilator RTL simulation, enables design-space exploration and architecture validation across hardware configurations.

DART makes three contributions:

\begin{enumerate}

\item \textbf{The first configurable hardware platform for dLLM inference.} DART comprises a systolic Transformer Engine with bidirectional FlashAttention support, a Vector-Scalar Sampling Engine, a unified dLLM-aware ISA, and a PyTorch-to-ISA compiler. The platform is validated through a tri-path simulation framework (analytical, transaction-level cycle-accurate, and accuracy simulators), cross-validated against AMD HBM2e and Verilator RTL simulation, and synthesized at 7nm (OpenROAD ASAP7~\cite{ASAP7}), providing area and power reference points.

\item \textbf{A hardware-friendly diffusion sampling engine.} Softmax are replaced with Stable-Max decomposition and top-$k$ selection are mapped into ISA primitives with phased reuse across a decoupled three-domain memory hierarchy (Vector/FP/Int SRAM), eliminating the mismatch between sampling workloads and GEMM-centric systems. A precision benchmark from FP64 to MXFP8 establishes that low precision preserves generation quality while reducing sampling to under 10\% of end-to-end latency.

\item \textbf{A novel dLLM-specific KV quantization with block-adaptive online smoothing (BAOS)}, it uses the warm-step recomputation of block-wise decoding as a zero-overhead calibration point, addressing channel-wise outliers whose statistics shift across denoising steps and enabling accurate MX-format KV quantization without offline calibration data.

\end{enumerate}

Using DART, we characterize the hardware behavior of the transformer forward pass and diffusion sampling stages, and sweep parameters in the design search space. DART achieves up to $\times$4.91 and $\times$2.06 TPS speedup over NVIDIA A6000 and H100 GPU, respectively, while delivering up to $\times$23.3 energy efficiency (tok/J) over A6000 and $\times$15.5 over H100. 
We find that BAOS is necessary to enable stable 4-bit KV cache quantization. With BAOS, a fully quantized configuration using 4-bit weights (MXINT4) and 8-bit activations (MXINT8) maintains accuracy on LLaDA-8B, matching or slightly exceeding the BF16 baseline (up to $+1.1$ percentage points on GSM8K).

\section{Background}
\label{se:Background}

\subsection{The dLLM Inference Pipeline}

\begin{figure}[]
    \centering
    \includegraphics[width=1.0\linewidth]{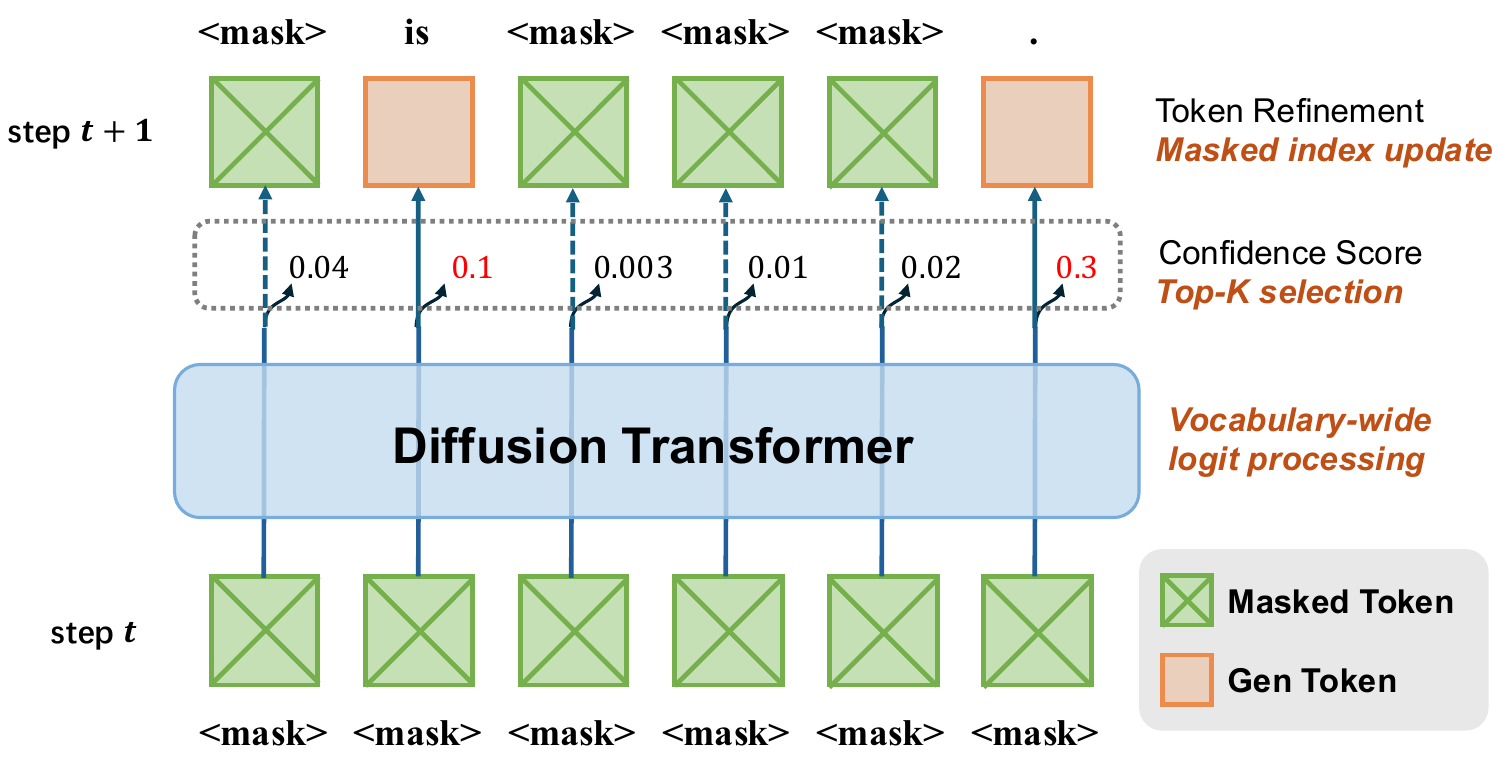}
    \caption{The dLLM inference pipeline. At each diffusion step, the transformer forward pass produces logits for all token positions; the sampling stage selects and commits the top-$k$ most confident tokens, progressively unmasking the sequence over $T$ steps.}
    \label{fg:dllm}
\end{figure}

Diffusion LLMs such as LLaDA~\cite{llada} and DREAM~\cite{ye2025dream} generate text through iterative masked denoising rather than autoregressive decoding. The inference pipeline alternates between two structurally distinct stages across $T$ diffusion steps, as illustrated in Fig.~\ref{fg:dllm}.

\textbf{Stage 1: Transformer forward pass.} At each diffusion step $t$, the full partially-masked sequence $x^{(t)} \in \mathbb{R}^{B \times L_\text{tot}}$ is processed by a standard transformer, producing vocabulary-wide logits $\mathbf{Z}^{(t)} \in \mathbb{R}^{B \times L_\text{tot} \times V}$, where $B$ is batch size, $L_\text{tot}$ total sequence length, and $V$ vocabulary size (up to $160\mathrm{k}$ in recent models). This stage is composed of GEMM-intensive attention and FFN layers and maps naturally onto systolic-array hardware.

A key structural difference from AR models is that dLLMs use \emph{bidirectional} attention: every position attends to every other position without a causal mask, producing a dense $L_\text{tot} \times L_\text{tot}$ attention matrix with no triangular sparsity to exploit. Consequently, every token's KV state may be invalidated when masked positions are updated between diffusion steps, creating a fundamentally different memory access pattern from AR decoding.

\textbf{Stage 2: Diffusion sampling.} The logits are converted to per-position probability distributions:
\begin{equation}
\mathbf{p}_i^{(t)} = \mathrm{softmax}\!\left(\mathbf{z}_i^{(t)}\right), \quad \mathbf{z}_i^{(t)} \in \mathbb{R}^{V}.
\end{equation}
A scalar confidence score $\max \mathbf{p}_i^{(t)}$ is computed per masked position; the top-$k$ positions with the highest confidence are unmasked and their predicted tokens committed. This stage involves vocabulary-wide reductions, rank-based sorting, and masked writes, operations with no analogue in AR decoding.

\begin{figure*}[h]
    \centering
    \includegraphics[width=0.9\linewidth]{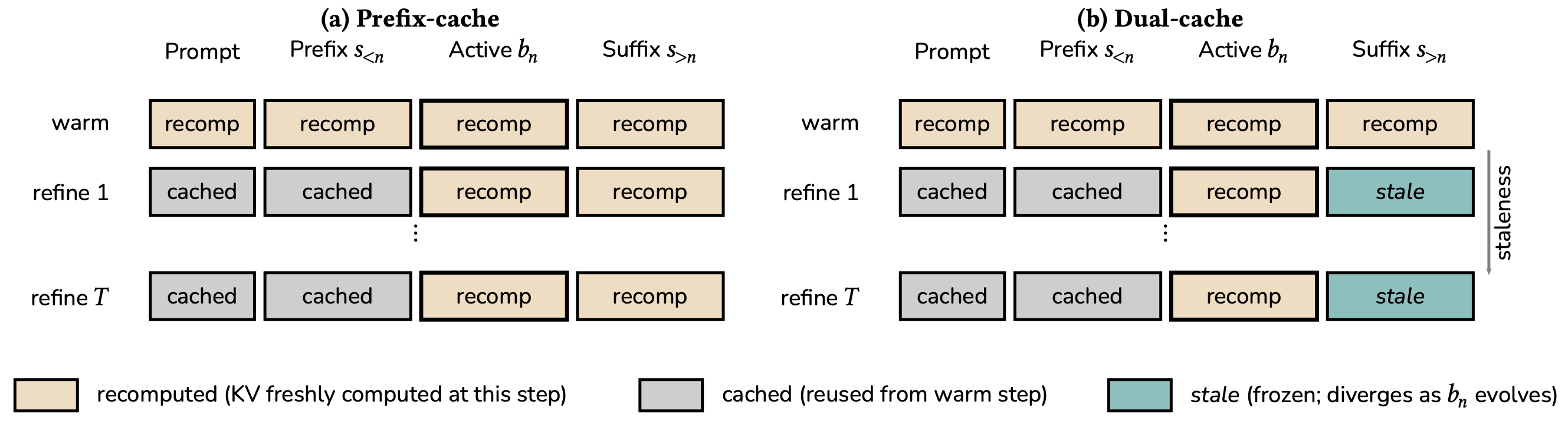}
    \caption{KV cache strategies in Fast-dLLM~\cite{wu2025fast}. Both modes begin with a warm step that caches KV. \textbf{(a)}~Prefix-cache truncates the cache to prefix-only after the warm step and recomputes active block and suffix KV at every refinement step (access range $x[s_n{:}]$), retaining full context freshness. \textbf{(b)}~Dual-cache retains the full warm-step cache; refinement steps input only the active block (access range $x[s_n{:}e_n)$) and replace its KV in-place each step. The suffix KV participates in attention but remains frozen from the warm step, introducing staleness that grows as active block tokens are progressively unmasked.}
    \label{fg:fast-dllm}
\end{figure*}

\subsection{Blocked Diffusion and KV Caching}
\label{subsec:kv_cache}

Block Diffusion~\cite{arriola2025block} first proposed partitioning the full generation sequence into blocks $\{b_1, b_2, \ldots, b_{N_B}\}$, each of length $L$ ($L_\text{tot} = L \times N_B$), where $N_B$ denotes the number of blocks. Generation proceeds autoregressively across blocks while diffusion denoising operates in parallel within each block over $T$ refinement steps. At each diffusion step, only the active subtensor $\mathbf{Z}^{(t)}_\text{active} \in \mathbb{R}^{B \times L \times V}$ is processed. In this original formulation, \emph{no KV cache is used}: the full KV states for all positions are recomputed from scratch at every diffusion step, making the transformer forward pass the dominant cost.

Fast-dLLM~\cite{wu2025fast} subsequently introduced KV caching into the blocked diffusion paradigm; we use Fast-dLLM as a representative example throughout this work. Both strategies begin with a \emph{warm step} that processes the full sequence (prompt, decoded prefix $s_{<n}$, active block $b_n$, and masked suffix $s_{>n}$) and caches KV for all positions. The two strategies differ in what is retained afterwards, as illustrated in Fig.~\ref{fg:fast-dllm}.

\textbf{Prefix-cache} truncates the cache to the prefix only after the warm step. Each refinement step reprocesses the sequence from the active block onward ($x^{(t)}[s_n{:}]$), temporarily recomputing active-block and suffix KV without caching them. This retains full context freshness at the cost of per-step active-block and suffix recomputation.

\textbf{Dual-cache} retains the full warm-step KV cache. Refinement steps process only the active block and replace its KV in-place each step, while suffix KV remains frozen from the warm step. This eliminates suffix recomputation but introduces a staleness gap that grows as active-block tokens are progressively unmasked.

The progression from no caching (Block Diffusion~\cite{arriola2025block}) to prefix-cache to dual-cache represents increasing approximation with correspondingly higher throughput, each creating distinct memory access patterns with no analogue in AR decoding's append-only KV buffer. As we show in Section~\ref{se:Quantization}, the warm step, shared by both caching modes, also provides a natural online calibration point for quantization-aware KV compression.

\section{DART Hardware Execution Model}
\label{se:Architecture}

\begin{figure}[t]
    \centering
    \includegraphics[width=1.0\linewidth]{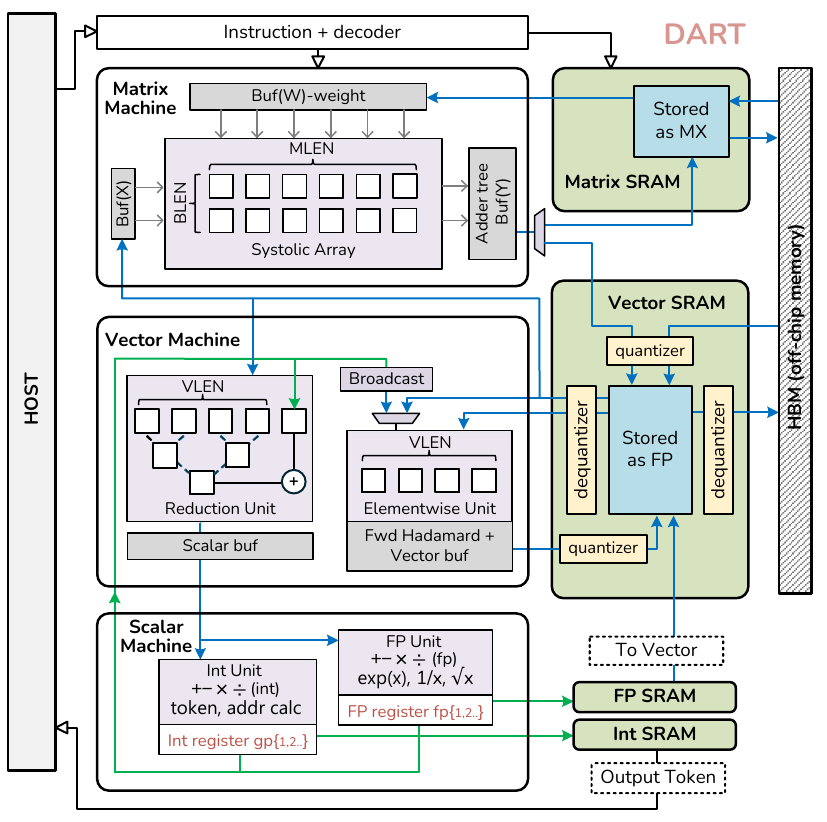}
    \caption{DART NPU Architecture Overview.}
    \label{fg:memory flow}
\end{figure}

DART is a configurable hardware platform for dLLM inference, covering the full execution pipeline: the GEMM-intensive transformer forward pass and the non-GEMM diffusion sampling stage. Figure~\ref{fg:memory flow} gives an overview of the architecture. The system adopts a multi-domain storage hierarchy to handle heterogeneous data types across the two stages: large-scale tensors (weights, KV cache, logits) reside in HBM using the low precision format; on-chip storage is partitioned into Vector SRAM, FP SRAM, and Int SRAM, each serving a distinct data path. The execution core comprises a Transformer Engine for GEMM-centric operations and a Vector-Scalar Engine for activation operations and non-GEMM sampling primitives, coordinated by an instruction decoder.

\subsection{Transformer Engine}
\label{subsec:transformer}

The DART Transformer Engine handles the GEMM-intensive forward pass of each dLLM diffusion step. It comprises three coordinated compute units (a Matrix Unit, a Vector Unit, and a Scalar Unit) and four on-chip SRAM banks. The Vector SRAM serves as a scratchpad for activations and intermediate results; the Matrix SRAM is dedicated to weights and KV tensors, with support for both transposed and non-transposed access patterns.

\subsubsection{Asymmetric Data Path}

To balance precision and efficiency, DART adopts an asymmetric precision policy across its memory and compute hierarchy. Activations are stored in the Vector SRAM in brain floating point (BF16). At the systolic array input boundary, BF16 activations are dynamically quantized to microscaling (MX) data format~\cite{rouhani2023microscaling}, producing per-block scale factors $s_a$; weights are loaded from Matrix SRAM in MX-format with precomputed per-block scales $s_w$. Inside each PE, each MX format product is first aligned to a common fixed-point domain by an arithmetic shift of $s_a + s_w$ (log-domain combination of the two block scales), and the shifted products are then accumulated into an INT32 register; the final INT32 value is cast to BF16 before being written back to the Vector SRAM. Weights and KV tensors are stored in HBM in the MX format (4/8-bit MXFP/MXINT) and loaded into the Matrix SRAM. Before writing newly computed KV vectors to the HBM cache, DART optionally applies Block-Adaptive Online Smoothing (Section~\ref{se:Quantization}) to suppress diffusion-specific channel-wise outliers prior to MX quantization with a calibration-free approach suited to block-wise diffusion decoding.

\subsubsection{Systolic Array}

Unlike AR decode, where each generation step produces a single new token per sequence ($M{=}B{\times}1$), every dLLM diffusion step processes the full token sequence. As shown in Fig.~\ref{fg:fast-dllm}, this occurs in two distinct phases per generation block: a \emph{warm step} that recomputes the KV cache over the entire context, and \emph{refinement steps} that operate on the active block alone. For a representative workload of $B{=}16$ sequences with block size $L{=}32$ and total context length $L_\text{tot}{=}256$, this gives $M_\text{warm}{=}16{\times}256{=}4096$ and $M_\text{refine}{=}16{\times}32{=}512$; the reduction dimension $K$ is the model hidden size, e.g.\ $K{=}4096$ for LLaDA-8B. Both $M$ values are far larger than in AR decode ($M{=}16$).

\begin{figure}[t]
    \centering
    \includegraphics[width=1.0\linewidth]{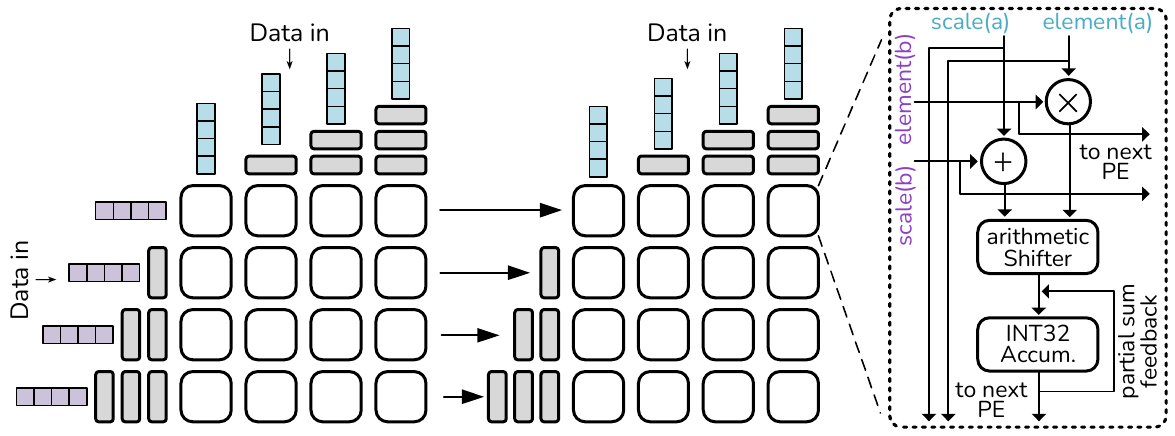}
    \caption{DART Systolic Array and PE Overview.}
    \label{fg:pe}
\end{figure}

This shapes the DART Matrix Unit design. Figure~\ref{fg:pe} illustrates the systolic array structure and the PE internal datapath. Each sub-array is a square $\texttt{BLEN}{\times}\texttt{BLEN}$ PE grid operating under output-stationary dataflow: partial sums remain stationary while operands stream in. Along the $K$ direction, $\texttt{MLEN}/\texttt{BLEN}$ sub-arrays are tiled side-by-side and fed a $\texttt{MLEN}$-wide slice of activation and weight in parallel; a result adder tree sums the partial sums via a single \texttt{M\_SUM} instruction. The full Matrix Unit replicates this structure as a grid, tiling rows and columns in steps of $\texttt{BLEN}$.

Full PE utilization therefore requires $\texttt{BLEN} \leq B{\times}L$; the warm step's larger $M$ is covered by additional tiles and imposes no tighter constraint. \texttt{MLEN} is sized for the $K$ dimension of projection GEMMs ($K = $ hidden size). Since head dimension $D \ll $ hidden size, \texttt{HLEN} $= \texttt{MLEN}/D$ attention heads are batched per call to fully utilize the \texttt{MLEN}-wide $K$ slice during attention computation. \texttt{VLEN} sets the lane width for elementwise and reduction operations in the Vector-Scalar Engine.

\subsubsection{Memory System and ISA}

All MX-format tensors stored in HBM have their scale factors and data blocks laid out contiguously per tensor to preserve memory alignment. Hardware prefetch engines integrated into both SRAMs enable background HBM-to-SRAM transfers, overlapping data movement with matrix and vector computations to sustain high bandwidth utilization.

The Transformer Engine is controlled by the DART ISA (Table~\ref{tab:DART_ISA}), which covers matrix operations (e.g., \texttt{M\_GEMM}, \texttt{M\_SUM}), vector elementwise and reduction operations, scalar FP primitives (e.g., softmax, layer norm, SiLU/GELU), and memory prefetch instructions. The ISA supports MHA, GQA, and MoE gating, covering the full range of dLLM transformer variants. The final transformer layer produces the logits tensor $\mathbf{Z}^{(t)} \in \mathbb{R}^{B \times L \times V}$, written to HBM and subsequently streamed into the Vector SRAM by the sampling pipeline described in Section~\ref{subsec:sampling}.

\begin{algorithm}[t]
\caption{DART Transformer Forward Pass (one diffusion step $t$)}
\footnotesize
\label{alg:transformer}
\begin{algorithmic}[1]
\Require $\texttt{model}$, $x \in \mathbb{R}^{B \times L}$, $\mathit{KV\_cache}$, $N_L$
\For{$\ell = 1$ \textbf{to} $N_L$}
  \State \blue{QKV projections} \hfill \green{MXINT8 act.\ $\times$ MXINT4 weights $\to$ INT32 $\to$ BF16 (see \S\ref{subsec:transformer})}
  \State $Q, K, V \gets \texttt{M\_GEMM}(x,\; W_Q^{(\ell)},\; W_K^{(\ell)},\; W_V^{(\ell)})$
  \State \blue{KV cache update} \hfill \green{BAOS quantization $\to$ HBM (block refresh)}
  \State $\hat{K}, \hat{V} \gets \texttt{BAOS}(K, V)$;\quad $\texttt{H\_STORE}(\mathit{KV\_cache}[\ell])$
  \State \blue{Attention} \hfill \green{Bidirectional FlashAttention (no causal mask)}
  \State $A \gets \texttt{FlashAttn}(Q,\; \mathit{KV\_cache}[\ell])$
  \State $x \gets \texttt{LayerNorm}(x + \texttt{M\_GEMM}(A,\; W_O^{(\ell)}))$
  \State \blue{FFN} \hfill \green{Dense or MoE gating}
  \State $x \gets \texttt{LayerNorm}(x + \texttt{FFN}(x,\; W_\mathit{FF}^{(\ell)}))$
\EndFor
\State $\mathbf{Z}^{(t)} \gets \texttt{M\_GEMM}(x,\; W_\mathit{lm}) \in \mathbb{R}^{B \times L \times V}$;\quad $\texttt{H\_STORE}(\mathbf{Z}^{(t)})$
\State \Return $\mathbf{Z}^{(t)}$
\end{algorithmic}
\end{algorithm}

\subsection{Diffusion Sampling Engine}
\label{subsec:sampling}

\begin{table}[t]
\centering
\small
\setlength{\tabcolsep}{4pt}
\renewcommand{\arraystretch}{1.1}
\caption{DART ISA overview.}
\label{tab:DART_ISA}
\begin{tabularx}{\columnwidth}{|l|X|c|}
\hline
\textbf{Class / Instruction} & \textbf{Description} & \textbf{\#} \\
\hline
\rowcolor{gray!15}\multicolumn{3}{|l|}{\textbf{Transformer instructions}} \\
\hline
Matrix (M)   & GEMM and GEMV operations, with or without matrix transposition & 6 \\
\hline
Vector (V)   & Element-wise and reduction operations, rotation for quantization & 16 \\
\hline
Scalar (S)   & Scalar INT and FP arithmetic (softmax, layer norm, SiLU/GELU) & 19 \\
\hline
HBM (H)      & Data transfers between HBM and the Matrix/Vector SRAMs & 3 \\
\hline
Control (C)  & Operation settings: HBM address, nested-loop configuration, branching & 8 \\
\hline
\rowcolor{gray!15}\multicolumn{3}{|l|}{\textbf{Sampling critical instructions}} \\
\hline
\texttt{V\_RED\_MAX\_IDX} & Vector reduction: finds maximum value and index in a single pass & — \\
\hline
\texttt{S\_ST\_FP}        & Store scalar FP value from FP register to FP SRAM & — \\
\hline
\texttt{S\_ST\_INT}       & Store scalar integer value from GP register to Int SRAM & — \\
\hline
\texttt{S\_MAP\_V\_FP}    & Transfer $L$ FP-format scalars from FP SRAM to Vector SRAM & — \\
\hline
\texttt{V\_TOPK\_MASK}    & Streaming top-$k$ sort; produces a vectorized boolean transfer mask & — \\
\hline
\texttt{V\_SELECT\_INT}   & Masked element-wise select on Int SRAM; equivalent to \texttt{torch.where()} & — \\
\hline
\end{tabularx}
\end{table}

\subsubsection{Hardware-Friendly Sampling Execution}

The sampling stage computes, per masked position, a confidence score over the full vocabulary logit vector $\mathbf{z} \in \mathbb{R}^V$ and selects the top-$k$ highest-confidence tokens. The standard approach computes the full softmax probability vector and indexes it at the argmax position:
\begin{equation}
i^* = \arg\max_i z_i, \qquad x_0\text{\_p} = \frac{e^{z_{i^*}}}{\sum_j e^{z_j}}.
\label{eq:softmax_standard}
\end{equation}
This requires materializing the full probability vector in SRAM before the confidence value can be extracted. DART instead applies the \emph{Stable-Max} reformulation. Letting $m = \max_i z_i$, the confidence simplifies to:
\begin{equation}
x_0\text{\_p} = \frac{e^{z_{i^*} - m}}{\sum_j e^{z_j - m}} = \frac{1}{\sum_j e^{z_j - m}},
\label{eq:stablemax}
\end{equation}
since $z_{i^*} = m$ by definition, making the numerator $e^0 = 1$. This decomposes into four sequential primitives that each map to a dedicated hardware unit: (1)~\texttt{V\_RED\_MAX\_IDX} computes $m$ and $i^*$ in one pass; (2)~\texttt{V\_EXP\_V} computes $e^{z_i - m}$ in-place, overwriting the logit buffer and eliminating a separate probability buffer; (3)~\texttt{V\_RED\_SUM} accumulates $\sum_j e^{z_j-m}$; and (4)~\texttt{S\_RECIP} takes the reciprocal to yield $x_0\text{\_p}$. No global synchronization is required between passes.

The full flow executes in four hardware-visible phases coordinated by the sampling ISA extensions (Table~\ref{tab:DART_ISA}). \textbf{Phase~1}(\emph{HBM $\to$ Vector $\to$ Scalar}): logit chunks are streamed from HBM into Vector SRAM via \texttt{H\_PREFETCH\_V}; Stable-Max and \texttt{V\_RED\_MAX\_IDX} compute per-position confidence scalars and argmax indices in a single pass. \textbf{Phase~2} (\emph{Scalar write-back}): confidence values and token indices are written to FP SRAM and Int SRAM, respectively, via \texttt{S\_ST\_FP}/\texttt{S\_ST\_INT}, decoupling the two scalar domains and preventing alignment conflicts. \textbf{Phase~3} (\emph{Scalar $\to$ Vector $\to$ Scalar}): the $L$ scalar confidence values are reconstructed into a dense vector via \texttt{S\_MAP\_V\_FP}; a streaming insertion-based Top-$k$ comparator (\texttt{V\_TOPK\_MASK}, $O(k)$ area) produces a boolean transfer mask. \textbf{Phase~4} (\emph{Integer masked update}): the Int Unit performs masked selection over Int SRAM token indices via \texttt{V\_SELECT\_INT}, committing the top-$k$ tokens to the output sequence. Algorithm~\ref{alg:asm_high_level} shows the complete hardware-aware intra-block sampling flow.

\begin{algorithm}[t]
\caption{DART Hardware-Aware Intra-Block Sampling}
\small
\label{alg:asm_high_level}
\begin{algorithmic}[1]
\small
\Require $\texttt{model}$, $prompt$, $T, B, L, V, V_\text{chunk}, VLEN$
\State $R \gets \lceil V / V_\text{chunk} \rceil$ \hfill \green{number of vocabulary chunks}
\State $x \in \mathbb{R}^{B \times L} \gets \texttt{blocked}[prompt, \mathit{mask\_id}]$
\State $k \in \mathbb{R}^{B} \gets \texttt{get\_num\_transfer\_tokens}(T)$
\For{$t = 1$ \textbf{to} $T$}
  \State $\mathbf{Z}^{(t)} \in \mathbb{R}^{B \times L \times V} \gets \texttt{Alg.~\ref{alg:transformer}}(x)$ \hfill \green{Transformer forward pass}
  \State $m\_idx \in \mathbb{R}^{B \times L} \gets (x == \mathit{mask\_id})$
  \For{$b = 1$ \textbf{to} $B$}
    \For{$l = 1$ \textbf{to} $L$}
      \State \blue{Phase~\ding{182}} \hfill \green{HBM $\rightarrow$ Vector $\rightarrow$ Scalar}
      \For{$r = 1$ \textbf{to} $R$}
        \State Preload $V_\text{chunk}$ from HBM to Vector SRAM
        \For{$i = 1$ \textbf{to} $(V_\text{chunk}/VLEN)$}
          \State $x_0\mathit{\_p\_scalar} \gets \texttt{Stable-Max}(\mathbf{Z}^{(t)})$
          \State $x_0\mathit{\_scalar} \gets \texttt{V\_RED\_MAX\_IDX}(\mathbf{Z}^{(t)})$
        \EndFor
      \EndFor
      \State \blue{Phase~\ding{183}} \hfill \green{Scalar write-back}
      \State $\mathit{SRAM}_\text{FP}[l] \gets x_0\mathit{\_p\_scalar}$ \quad via \texttt{S\_ST\_FP}
      \State $\mathit{SRAM}_\text{Int}[l] \gets x_0\mathit{\_scalar}$ \quad via \texttt{S\_ST\_INT}
    \EndFor
    \State \blue{Phase~\ding{184}} \hfill \green{Scalar (FP) $\rightarrow$ Vector $\rightarrow$ Scalar (Int)}
    \State $x_0\mathit{\_p} \gets \texttt{S\_MAP\_V\_FP}(\mathit{SRAM}_\text{FP}[1{:}L])$
    \State $\mathit{transfer\_idx} \gets \texttt{V\_TOPK\_MASK}(x_0\mathit{\_p},\, m\_idx,\, k)$
    \State \blue{Phase~\ding{185}} \hfill \green{Integer masked update}
    \State $x_0 \gets \mathit{SRAM}_\text{Int}[1{:}L]$
    \State $x_0 \gets \texttt{V\_SELECT\_INT}(m\_idx,\, x_0,\, x)$
    \State $x \gets \texttt{V\_SELECT\_INT}(\mathit{transfer\_idx},\, x,\, x_0)$
  \EndFor
\EndFor
\State \Return $x$
\end{algorithmic}
\vspace{2pt}
\begin{flushleft}
\footnotesize{$V_\text{chunk}$ controls tiling granularity: $V_\text{chunk} < V$ enables edge-device operation with minimal SRAM; $V_\text{chunk} = V$ maximizes data reuse on resource-rich platforms.}
\end{flushleft}
\end{algorithm}

\subsubsection{Decoupled Mixed-Precision Memory Hierarchy}

To prevent address-decoder contention and control-path interference between the transformer and sampling stages, DART decouples on-chip storage into three physically isolated domains:

Vector SRAM serves as the primary high-throughput data path. During transformer execution, it holds tiled activation; during sampling, it buffers logit chunks streamed from HBM and intermediate $\mathit{exp\_shifted}$ values computed in-place. Its footprint is governed by:
\begin{align}
\text{Vector elements} &=
\begin{cases}
3 \cdot B \cdot L + V_\text{chunk}, & V_\text{chunk} < V\\
3 \cdot B \cdot L + V \cdot L \cdot R, & \text{otherwise}
\end{cases} \label{eq:vector}
\end{align}

FP SRAM stores per-position scalar confidence values ($L$ entries of BF16) and intermediate FP scalars from transcendental computations. It interfaces with the FP Unit through a dedicated register file and with the Vector SRAM through \texttt{S\_MAP\_V\_FP}.
\begin{equation}
\text{FP elements} = \max\{L,\, VLEN\} \label{eq:fp}
\end{equation}

Int SRAM holds integer token indices and boolean masks, physically isolated from the vector data path and coupled to the host output interface via a FIFO buffer to deliver final token IDs.
\begin{equation}
\text{Int elements} = 2 \cdot B \cdot L \label{eq:int}
\end{equation}

\begin{figure}[t]
    \centering
    \includegraphics[width=1.0\linewidth]{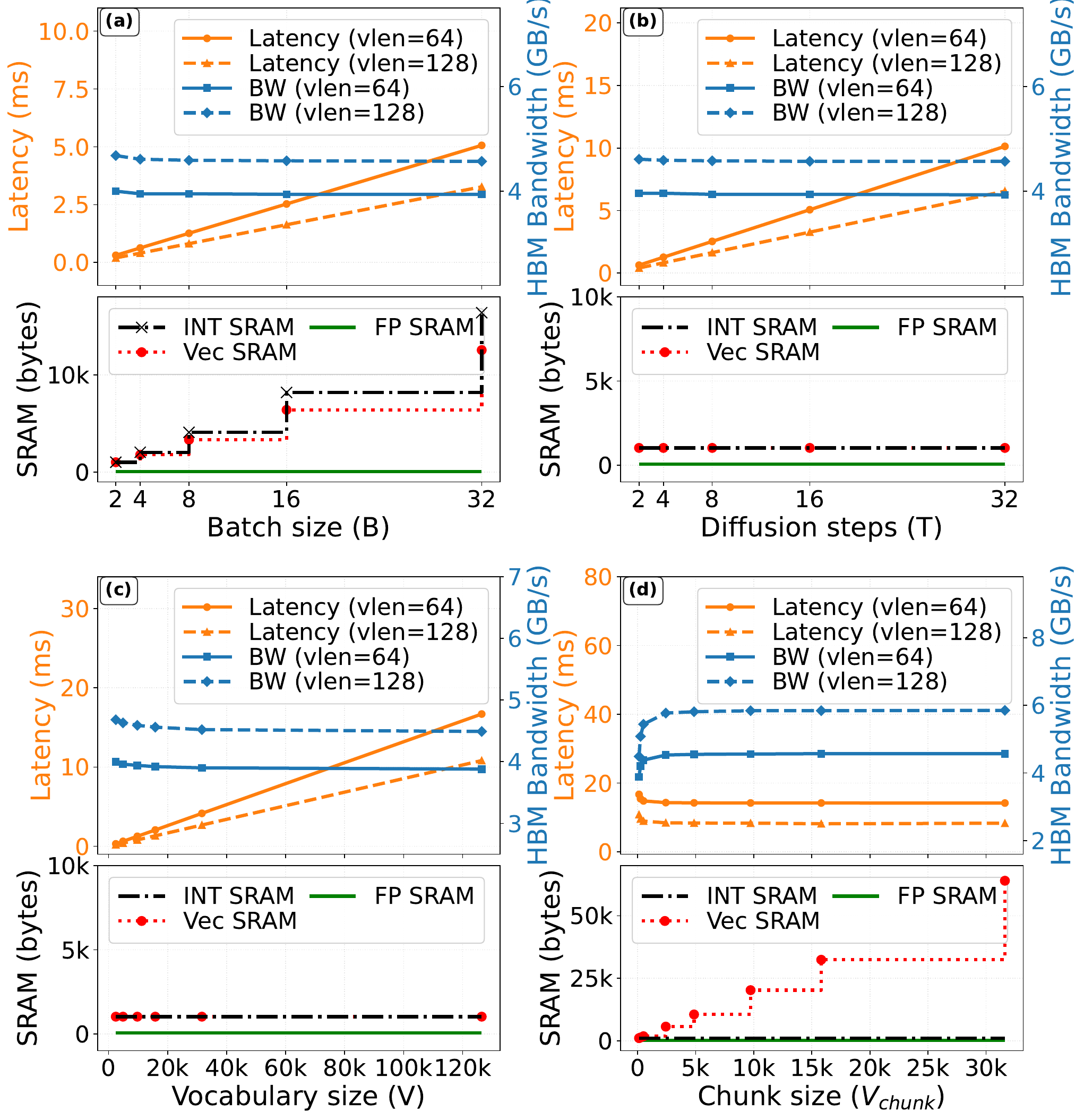}
    \caption{Sampling engine latency and on-chip memory utilization under parameter sweeps: (a) batch size $B$, (b) diffusion steps $T$, (c) vocabulary size $V$, (d) chunk size $V_\text{chunk}$.}
    \label{fg:latency_memory}
\end{figure}

Figure~\ref{fg:latency_memory} characterizes the performance of the proposed diffusion-sampling primitives under systematic sweeps of key workload parameters, with the \texttt{model()} execution excluded to isolate sampling overhead. All experiments fix the generation length to $L=64$ and $VLEN{=}64,128$. 

We first evaluate scalability with respect to batch size and diffusion steps by sweeping $B, T \in \{2,4,8,16,32\}$ while fixing $V=2\mathrm{k}$ and $V_{chunk}=128$, as shown in Fig.~\ref{fg:latency_memory}(a) and (b), respectively.
We then study the impact of vocabulary size by sweeping $V$ from 2$\mathrm{k}$ to 128$\mathrm{k}$ under a fixed configuration of $B=2$, $T=1$, and $V_{chunk}=128$, as reported in Fig.~\ref{fg:latency_memory}(c).
Finally, Fig.~\ref{fg:latency_memory}(d) examines the effect of chunk size $V_{chunk}$ by sweeping it from 128 to 30$\mathrm{k}$ at the largest vocabulary setting ($V=128\mathrm{k}$), with $B=2$ and $T=1$ fixed.

As shown in Fig.~\ref{fg:latency_memory}(a)–(c), the sampling latency scales approximately linearly with $B$, $T$, and $V$, while the achieved HBM bandwidth remains nearly constant. This indicates stable bandwidth utilization and predictable performance scaling across these workload dimensions. In contrast, Fig.~\ref{fg:latency_memory}(d) shows that increasing the chunk size $V_{chunk}$ reduces latency and improves effective HBM bandwidth by amortizing control and reduction overheads. Beyond approximately 4$\mathrm{k}$ entries, both metrics saturate, suggesting that large Vector SRAM capacities are not required to achieve near-peak efficiency, which is favorable for edge-oriented deployments.

The bottom inset of each sub-figure reports the on-chip SRAM footprint (in bytes) for the $VLEN=64$ configuration. The SRAM cost is evaluated as the number of elements determined by Eq.~\ref{eq:vector}--Eq.~\ref{eq:int}, multiplied by their respective byte widths. For the vector SRAM, $V_{\text{chunk}} < V$ corresponds to the edge mode, whereas in performance mode, one can preload $V \cdot L \cdot R$ elements from HBM into SRAM with $R < B$ and $R$ divisible by $B$. We report SRAM usage only for $VLEN=64$ because the Int SRAM and Vector SRAM footprints are invariant to $VLEN$, and the FP SRAM differs only by a small additive term proportional to $VLEN$. Under the proposed primitive design and across all parameter sweeps in the edge scenario, the on-chip SRAM footprint is dominated by the batch size $B$ and $V_{chunk}$; varying $T$, $V$, or $VLEN$ does not materially affect the overall requirement.

\section{Simulation Framework and KV Quantization}
\label{se:Simulation}

\subsection{Analytical Simulator}
\label{subsec:analytical}

The analytical simulator provides closed-form latency, power, and energy estimates for rapid design-space exploration across hardware configurations (\texttt{BLEN}, \texttt{MLEN}, \texttt{VLEN}, \texttt{HLEN}, SRAM capacity, HBM) and workload parameters (model architecture, inference paradigm, prompt/generation length, denoising steps, block size).

Each ISA instruction maps to a pipelined cycle count from a hardware-derived latency library. $T_\text{cmp}$ is the compute-bound cost under infinite memory bandwidth. The memory cost, $T_\text{mem}$, is derived from the effective bandwidth of two physically independent SRAM paths: Matrix SRAM (weights, KV) and Vector SRAM (activations), each bounded by the on-chip port bandwidth and the off-chip HBM specification (stack count and per-stack bandwidth). With the two paths being accessed concurrently, per-operation latency applies a roofline model, $T_\text{op} = \max(T_\text{cmp},\, T_\text{mem})$, at the instruction granularity across all dLLM operators.

For block-diffusion paradigms, the simulator switches memory strategy per phase: the warm step uses strategy: ($M = B{\times}L_\text{tot}$, weights streamed) and each refinement step uses strategy: ($M = B{\times}L$, KV resident in SRAM), with per-block latency $T_\text{block} = T_\text{warm}(L_\text{tot}) + (\texttt{steps}{-}1)\cdot T_\text{refine}(L)$. The sampling stage models diffusion denoising over $\mathbf{Z} \in \mathbb{R}^{B \times L \times V}$, each applying the same roofline; when the logit footprint exceeds Vector SRAM, bandwidth degrades to $\min(BW_\text{HBM},\, BW_\text{VSRAM})$.

Power and area are estimated from parametric models calibrated against DART RTL post-synthesis results at the Synopsis DC tool with a library of 7\,nm/1\, GHz (ASAP7~\cite{ASAP7}).

\subsection{Rust-Based Cycle-Accurate Simulator}
\label{subsec:simulator}

The DART simulator is implemented in Rust and built on top of Ramulator~\cite{Ramulator}, which provides a detailed HBM DRAM model including bank-level parallelism, row-buffer policies, and refresh overhead. All components, from individual ISA instructions to compound sequences such as the full Transformer Engine and Sampling Engine, are exercised by running DART compiler-generated assembly, with functional correctness cross-checked against PyTorch baselines. The simulator decodes DART ISA instructions (Table~\ref{tab:DART_ISA}), models in-order issue with stall-on-dependency, and reports cycle-accurate latency, effective HBM bandwidth, and on-chip SRAM utilization. These outputs are cross-validated in Section~\ref{se:Validation}.

\subsection{Weight and Activation Quantization}
\label{se:WAQuantization}
DART is the first to adapt microscaling data formats to block-wise dLLM quantization, building on the PTQ framework established in PLENA~\cite{plena}. We adopt MXINT4 as the target weight format, as MXINT consistently outperforms MXFP under aggressive 4-bit weight quantization~\cite{plena}. 

For weight quantization, we apply the output-norm guided blockwise clipping method from PLENA's accuracy simulator~\cite{plena}, which integrates a clipping search into GPTQ's~\cite{gptq} iterative Hessian-based error propagation flow. Given a linear layer $\mathbf{Y} = \mathbf{X}\mathbf{W}^\top$ with weights $\mathbf{W} \in \mathbb{R}^{N \times K}$ and calibration inputs $\mathbf{X} \in \mathbb{R}^{M \times K}$, GPTQ processes the weights in column-wise blocks of size $B$ (aligned with the MX block size). For each row within a block, a clipping percentile $p \in \mathcal{P} \subset [0.5, 0.99]$ shrinks the effective representable range to $[p\min_w,\, p\max_w]$, trading off clipping error on outliers against finer resolution for inliers. Rather than selecting percentiles to minimize weight reconstruction error, the search minimizes the output reconstruction error: 

\begin{equation} 
\label{eq:clip_search}
P^\star_b = \arg\min_{P_b \in \mathcal{P}^N} \left\| \mathbf{X}_b \left( \mathbf{W}_b - Q(\mathbf{W}_b;\, P_b, \tau) \right)^\top \right\|_2^2, 
\end{equation} 

where $\mathbf{W}_b \in \mathbb{R}^{N \times B}$ and $\mathbf{X}_b \in \mathbb{R}^{M \times B}$ are the weight and activation slices for block $b$, $P_b = (p_1, \dots, p_N) \in \mathcal{P}^N$ collects the per-row clipping percentiles, $\tau = (d, n, B)$ specifies the MX data format (datatype $d$, bit-width $n$, block size $B$), and $Q(\cdot;\, P_b, \tau)$ denotes per-row MX-format quantization under the clipped range. After quantizing each block with $P_b^\star$, GPTQ compensates the resulting error across all remaining blocks via the Hessian update. In Tables~\ref{tab:v1_quant}, we evaluate clipping search and GPTQ both independently and in combination: \texttt{x-clip} denotes weight-norm guided clipping search and \texttt{y-clip} denotes output-norm guided clipping search (Equation~\ref{eq:clip_search}). We additionally adapt QuaRot~\cite{quarot} to blocked dllm inference as baseline.

For activations, DART stores BF16 in the Vector SRAM and dynamically quantizes to MXINT8 at the systolic array boundary (Section~\ref{subsec:transformer}), avoiding the accuracy loss of static low-bit activation quantization while enabling efficient integer MACs.

\subsection{Block-Adaptive Online Smoothing for KV Quantization}
\label{se:Quantization}

KV activations in transformer-based AR LLMs are known to exhibit channel-wise outliers~\cite{smoothquant}. We profile this phenomenon on LLaDA-8B-Instruct under the GSM8K benchmark using the dual-cache block-diffusion paradigm (128 denoising steps, generation length 128, block length 32, giving 4 blocks of 32 steps each), collecting per-channel KV activation statistics across all $N_L$ layers at every diffusion step. Diffusion decoding exhibits a similar phenomenon under block-wise KV caching; a small fraction of KV channels displays magnitudes $13$--$19\times$ larger than the global mean, producing severe distribution imbalance. 

\subsubsection{Warm-Step Calibration}

In Fast-dLLM-style decoders~\cite{wu2025fast}, the \emph{warm step} at the start of each generation block recomputes the full KV cache for the entire sequence, as described in Section~\ref{se:Background}. Under the same profiling setup, we further find that within an active generation block, the dominant outlier channels identified during the warm step remain largely consistent across subsequent refinement steps: over $70\%$ of the top-$K_\text{out}$ outlier channel indices are shared between the warm step and all subsequent 128 refinement steps across all layers. This stability means that scaling factors computed once during the warm step can be reused throughout the refinement steps with limited mismatch, making online calibration practical with negligible overhead.

\begin{figure}[t]
    \centering
    \includegraphics[width=0.8\linewidth]{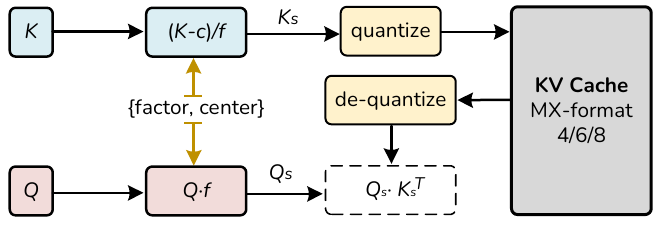}
    \caption{Integration of Block-Adaptive Online Smoothing. }
    \label{fg:system}
\end{figure}

\subsubsection{Method}

Let the key (or value) tensor be $x \in \mathbb{R}^{B \times H \times S \times D}$, where $B$ is batch size, $H$ is the number of attention heads, $S$ is the sequence length within the active block, and $D$ is the head dimension. For each generation block, we treat the warm step as an online calibration point and compute a per-channel scaling factor of shape $(B, H, 1, D)$ by reducing over the sequence dimension $S$.

In the \emph{mean}-centered variant, the center $c$ is the temporal mean and the scaling factor is:
\begin{equation}
f = \max(x_{\max} - c,\ c - x_{\min}),
\end{equation}
where $x_{\max}$ and $x_{\min}$ denote per-channel temporal extrema. In the \emph{minmax} variant, the center is the midpoint between $x_{\min}$ and $x_{\max}$, with the same symmetric radius.

To further suppress extreme channels without over-amplifying weak ones, we apply a per-channel power transform:
\begin{equation}
f \leftarrow f^{\alpha}, \quad \alpha \in [0, 1.0].
\end{equation}
This compresses the dynamic range of scaling factors: outlier-dominated channels are damped while low-magnitude channels are moderately inflated. The choice of $\alpha$ provides a practical DSE knob: Section~\ref{subsec:kv_quant} ablates its effect across workloads and datasets, showing that the optimal value varies with model and task. Before writing a KV tensor to the cache, we apply the asymmetric normalization $(x - c) / f$ and feed the result into the MXINT/MXFP block quantizer, which now operates on a smoothed distribution with flattened per-channel dynamic range.

\subsubsection{System Integration}

Figure~\ref{fg:system} illustrates the integration of BAOS into the KV cache pipeline. At the warm step, per-channel scaling factors $f$ are computed, and the normalized key tensors $K_s = (K - c)/f$ are written to the cache. During refinement attention, rather than unscaling $K_s$ before the multiply (which requires reading the full KV cache), we instead apply the inverse scaling to the query: $Q_s = Q \cdot f$, fusing this directly into the $Q_s K_s^\top$ computation. In long-prefix decoding scenarios where the number of query tokens is far smaller than the number of cached key tokens, this avoids a bandwidth-intensive pass over the entire KV cache and keeps the effective memory overhead negligible.

\section{Simulator Cross-Validation}
\label{se:Validation}
To validate the DART simulator, we adopt a \emph{component-wise cross-validation} strategy in this section: the memory subsystem model is validated against physical HBM2e measurements, and the compute pipeline model is validated against Verilator RTL simulation.

\subsection{Memory Subsystem Validation}
\label{subsec:mem_validation}

To validate the memory subsystem model of the DART cycle-accurate simulator (Section~\ref{subsec:simulator}), we collect physical HBM2e measurements from an \textbf{AMD Alveo V80 FPGA} (2 HBM2e stacks, 64 pseudo-channels, datasheet peak 819\,GB/s). To isolate HBM2e throughput from host-interface effects, we design a custom AXI bandwidth benchmark that runs entirely within the FPGA fabric: an AXI master issues transactions directly to the HBM2e controller without traversing PCIe. The AXI master is configured with 256-bit data width (32\,B per beat), burst length 128 (4\,KB per burst), 3 outstanding write transactions, and 4 outstanding read transactions. The DART simulator is configured with 2 stacks to match the physical FPGA, and separately with 4 stacks to project the peak bandwidth of our target NPU configuration. In the DART simulator, we evaluate the read and write bandwidth by measuring the average throughput during continuous read and write operations of 64 MB of data within the same memory region.

\begin{table}[t]
\centering
\caption{Memory subsystem validation: DART simulator vs.\ physical HBM2e on AMD Alveo V80 (2-stack, 64 ch, datasheet peak 819\,GB/s). 4-stack projects peak NPU performance (no physical counterpart).}
\label{tab:mem_validation}
\renewcommand{\arraystretch}{1.1}
\footnotesize
\begin{tabularx}{\columnwidth}{|X|c|c|}
\hline
\textbf{Metric} & \textbf{Write} & \textbf{Read} \\
\hline
\rowcolor{gray!15}\multicolumn{3}{|l|}{\textbf{2-stack (64 ch): cross-validation}} \\
\hline
Datasheet spec (GB/s)           & \multicolumn{2}{c|}{819} \\
Physical BW (GB/s)              & 763 (93\%) & 705 (86\%) \\
DART sim BW (GB/s)              & 862.5 & 846.4 \\
Sim error vs.\ physical         & $+$13.0\% & $+$20.1\% \\
Sim error vs.\ spec             & $+$5.3\%  & $+$3.3\% \\
\hline
\rowcolor{gray!15}\multicolumn{3}{|l|}{\textbf{4-stack (128 ch): peak NPU performance}} \\
\hline
DART sim BW (GB/s)              & 1739.1 & 1415.9 \\
\hline
\end{tabularx}
\end{table}

Under write-only and read-only traffic the physical measurements reach 93\% and 86\% of the datasheet peak, with the gap attributable to AXI scheduling overhead and bank-conflict penalties not captured by the specification. The DART simulator errs by only $+$5.3\% (write) and $+$3.3\% (read) against the datasheet, indicating that the HBM2e model in the simulator accurately captures the specification. The larger error against physical measurements ($+$13.0\% write, $+$20.1\% read) reflects the simulator modeling ideal bank-level parallelism, whereas the physical device incurs contention and refresh overhead under sustained traffic. The 4-stack projection demonstrates the peak bandwidth achievable by our target NPU configuration.

\subsection{Compute Pipeline Validation}
\label{subsec:compute_validation}

We validate the DART compute pipeline bottom-up against Verilator RTL simulation: first at the single instruction level to characterize per-unit accuracy, then at the compound sequence level to show how individual errors propagate to realistic workloads. Vector unit RTL cycles are measured at the top-level vector machine wrapper; GEMM and FlashAttention cycles are measured at the matrix engine top-level. Table~\ref{tab:compute_validation} presents results in a unified view.

\begin{table}[t]
\centering
\caption{Compute pipeline validation: DART simulator vs.\ Verilator RTL ($VLEN=8$, $BLEN=4$).}
\label{tab:compute_validation}
\renewcommand{\arraystretch}{1.1}
\footnotesize
\begin{tabularx}{\columnwidth}{|X|>{\centering\arraybackslash}p{30pt}|>{\centering\arraybackslash}p{28pt}|>{\raggedleft\arraybackslash}p{16pt}|}
\hline
\textbf{Primitive / Sequence                      } & \textbf{RTL (cyc)} & \textbf{Sim (cyc)} & \textbf{Error} \\
\hline
\rowcolor{gray!15}
\multicolumn{4}{|l|}{\textbf{Single instructions} (pipeline RTL-calibrated; Sim $\equiv$ RTL by construction)} \\
\hline
\texttt{V\_ADD\_VV} & \multicolumn{2}{c|}{7}  & 0\% \\
\texttt{V\_EXP\_V} & \multicolumn{2}{c|}{7}  & 0\% \\
\texttt{V\_RED\_MAX} & \multicolumn{2}{c|}{4}  & 0\% \\
\texttt{V\_RED\_SUM} & \multicolumn{2}{c|}{20} & 0\% \\
\texttt{V\_TOPK\_MASK} ($L=32, k=8$) & \multicolumn{2}{c|}{32} & 0\% \\
\texttt{V\_TOPK\_MASK} ($L=64, k=16$) & \multicolumn{2}{c|}{64} & 0\% \\
\hline
\rowcolor{gray!15}
\multicolumn{4}{|l|}{\textbf{Compound sequences}} \\
\hline
Softmax  & 43 & 38 & $-11.6\%$ \\
GEMM $[1 \times 64 \times 64]$ (proj., 16 tiles) & 86 & 80 &  $-7.0\%$ \\
FlashAttention ($d=64$, $H=2$, 6 GEMMs) & 401 & 365 & $-8.9\%$ \\
\hline
\rowcolor{gray!15}
\multicolumn{3}{|l|}{\textbf{$\hookrightarrow$ FlashAttention layer per-op breakdown}} & \textbf{Error} \\
\hline

$\blacktriangleright$ Q-projection$(1\times64)@(64\times64)$, 16 tiles &  86 &  80 &  $-6$ \\
$\blacktriangleright$ K-projection$(1\times64) @ (64\times64)$, 16 tiles &  86 &  80 &  $-6$ \\
$\blacktriangleright$ V-projection$(1\times64) @ (64\times64)$, 16 tiles &  86 &  80 &  $-6$ \\
$\blacktriangleright$ $QK^\top$$(1\times32) @ (32\times1)$, $\times 2$ heads, 1 tile &  11 &   5 &  $-6$ \\
$\blacktriangleright$ $AV$$(1\times1) @ (1\times32)$, $\times 2$ heads, 8 tiles &  46 &  40 &  $-6$ \\
$\blacktriangleright$ O-projection$(1\times64) @ (64\times64)$, 16 tiles &  86 &  80 &  $-6$ \\
\hline
\end{tabularx}
\end{table}

Per-instruction cycle counts are measured directly from Verilator RTL simulation and used to populate the simulator's latency library; single-instruction error is therefore zero by construction. The RTL characterises each unit's pipelined throughput.

Since single-instruction latencies are exact by construction, all compound-sequence error originates from pipeline inter-stage costs that the simulator does not model. For GEMM, the simulator accumulates pipelined throughput ($1{+}\texttt{BLEN}{=}5$ cycles/tile) without modelling the first-tile pipeline-fill overhead of ${\approx}6$ cycles. The per-op breakdown confirms this: every operation incurs a constant $-6$\,cycle error regardless of tile count or sequence length, demonstrating that the error is a fixed structural overhead rather than a function of workload size. At production projection dimensions ($N_\text{tiles}{=}16$), this corresponds to $-$7\%; at larger tile counts the relative impact diminishes further. For Softmax, the $-$5\,cycle error similarly reflects unmodelled pipeline-drain overhead between the sequential reduction and elementwise stages. Including these pipeline overheads will be considered as future work.

Beyond compute pipeline validation, we cross-validate the two simulators on the diffusion sampling stage. Table~\ref{tab:sampling_validation} reports results for a sampling block with $T{=}1$, $B{=}16$, $L{=}32$, and $V{=}126\mathrm{k}$, using $R{=}1$ (full block logits preloaded into Vector SRAM per iteration to balance on-chip capacity against HBM transfers), at $VLEN{=}2048$. The transactional and analytical simulators agree to within 4\%, consistent with the compound-sequence errors characterized above, while the analytical simulator provides a ${\sim}120{\times}$ wall-clock speedup, making it the practical tool for the design-space sweeps reported in Section~\ref{se:Results}.

\begin{table}[t]
\centering
\caption{Cross-validation of DART transactional and analytical simulators on a sampling block.}
\small
\begin{tabular}{llc}
\toprule
\textbf{Evaluator} & \textbf{Simulated Time}  & \textbf{Run Time} \\
\midrule
DART transactional  & 0.99 $\mathrm{ms}$ & 20 mins$^\dagger$ \\
DART analytic       & 0.95 $\mathrm{ms}$ ($-$4.0\%) & <10 s \\
\bottomrule
\multicolumn{3}{p{0.9\columnwidth}}{\footnotesize $^\dagger$ Includes ASM code generation; actual simulator execution is 3--4 min. The remainder is ASM file I/O overhead.}
\end{tabular}
\label{tab:sampling_validation}
\end{table}
\section{Experimental Results}
\label{se:Results}

\subsection{Quantization Quality}
\label{subsec:kv_quant}

We evaluate generation quality under MXINT4 KV-cache and weight quantization using the DART accuracy simulator, across two cache structures under Fast-dLLM (LLaDA 8B backend, Table~\ref{tab:v1_quant}). For KV quantization we compare a BF16 baseline, naive KV4, rotation-based smoothing QuaRot~\cite{quarot}, and Block-Adaptive Online Smoothing (DART-BAOS) with mean ($\bar{\alpha}$) and minmax ($\hat{\alpha}$) calibration at $\alpha{\in}\{1.0,\,0.9,\,0.6\}$; for weight quantization we compare W4 (MXINT4) with GPTQ~\cite{gptq} calibration and activation clipping variants. KV cache and weight quantization are searched independently as two separate tracks, each evaluated in isolation to identify its best configuration, and the selected configurations (marked in green) are then combined with DART quantized sampling (BF16 and MXFP8) to form the final fully-quantized DART model, reported in the \textit{Full Quantization} rows at the bottom of each section. All tested models are configured in their GitHub code-base reference setup.

\begin{table}[t]
\centering
\caption{Quantization quality of LLaDA series.}
\label{tab:v1_quant}
\renewcommand{\arraystretch}{1.1}

\footnotesize
\setlength{\tabcolsep}{2pt}
\begin{tabularx}{\columnwidth}{|l|Xccc|}
\hline
\textbf{Cache} & \textbf{Configuration} & \shortstack{\textbf{GSM8K} \textbf{(0-shot)}} & \shortstack{\textbf{GSM8K} \textbf{(1-shot)}} & \textbf{HumanEval} \\
\hline
\multirow{19}{*}{\textbf{Prefix}}
 & Baseline~\cite{wu2025fast}$^\dagger$                 & 0.7172         & 0.7521         & 0.4268         \\
\cline{2-5}
 & \multicolumn{4}{l|}{\cellcolor{gray!12}\textit{Sampling Quantization}} \\
 & DART (BF16)                       & 0.7202         & 0.7513        & 0.4268         \\
 & DART (MXFP8)                      & 0.7218         & 0.7362         & 0.3719         \\
\cline{2-5}
 & \multicolumn{4}{l|}{\cellcolor{gray!12}\textit{KV cache Quantization}} \\
 & KV4                                       & 0.7187         & 0.7506         & 0.4268         \\
 & QuaRot~\cite{quarot}                      & 0.7316         & 0.7415         & 0.4024         \\
 & DART-BAOS ($\bar{\alpha}{=}1.0$)              & \best{0.7362}  & 0.7483         & 0.4207         \\
 & DART-BAOS ($\hat{\alpha}{=}1.0$)              & 0.7233         & \best{0.7521}  & 0.4207         \\
 & DART-BAOS ($\bar{\alpha}{=}0.9$)              & 0.7248         & 0.7346         & 0.4268         \\
 & DART-BAOS ($\hat{\alpha}{=}0.9$)              & 0.7180         & 0.7506         & 0.4268         \\
 & DART-BAOS ($\bar{\alpha}{=}0.6$)              & 0.7286         & 0.7437         & \best{0.4390}  \\
 & DART-BAOS ($\hat{\alpha}{=}0.6$)              & 0.7225         & 0.7513         & \best{0.4390}  \\
\cline{2-5}
 & \multicolumn{4}{l|}{\cellcolor{gray!12}\textit{Weight Quantization}} \\
 & W4                                        & \best{0.7180}  & 0.7286         & 0.4268         \\
 & DART x-clip(W4)                                & 0.7074         & \best{0.7483}  & \best{0.4329}  \\
\cline{2-5}
 & \multicolumn{4}{l|}{\cellcolor{gray!12}\textit{Full Quantization (KV4+W4+A8+S16)}} \\
 & DART            & 0.7286         & 0.7165         & 0.4268         \\
\hline
\multirow{19}{*}{\textbf{Dual}}
 & Baseline~\cite{wu2025fast}$^\dagger$                 & 0.7202         & 0.7331         & 0.3658         \\
\cline{2-5}
 & \multicolumn{4}{l|}{\cellcolor{gray!12}\textit{Sampling Quantization}} \\
 & DART (BF16)                       & 0.7218         & 0.7309         & 0.3658         \\
 & DART (MXFP8)                      & 0.6937         & 0.7400         & 0.4085         \\
\cline{2-5}
 & \multicolumn{4}{l|}{\cellcolor{gray!12}\textit{KV cache Quantization}} \\
 & KV4                                       & 0.7157         & 0.7415         & 0.3475         \\
 & QuaRot~\cite{quarot}                      & 0.7111         & 0.7354         & 0.3658         \\
 & DART-BAOS ($\bar{\alpha}{=}1.0$)              & 0.7013         & \best{0.7513}  & 0.3719         \\
 & DART-BAOS ($\hat{\alpha}{=}1.0$)              & 0.7058         & 0.7445         & \best{0.3780}  \\
 & DART-BAOS ($\bar{\alpha}{=}0.9$)              & 0.7142         & 0.7316         & 0.3536         \\
 & DART-BAOS ($\hat{\alpha}{=}0.9$)              & 0.7165         & 0.7293         & 0.3536         \\
 & DART-BAOS ($\bar{\alpha}{=}0.6$)              & \best{0.7218}  & 0.7407         & 0.3719         \\
 & DART-BAOS ($\hat{\alpha}{=}0.6$)              & \best{0.7218}  & 0.7354         & 0.3719         \\
\cline{2-5}
 & \multicolumn{4}{l|}{\cellcolor{gray!12}\textit{Weight Quantization}} \\
 & W4                                        & \best{0.7074}  & 0.7149         & \best{0.3231}  \\
 & DART x-clip(W4)                                & 0.6907         & \best{0.7346}  & \best{0.3231}  \\
\cline{2-5}
 & \multicolumn{4}{l|}{\cellcolor{gray!12}\textit{Full Quantization (KV4+W4+A8+S16)}} \\
 & DART            & 0.6998         & 0.7415         & 0.3415         \\
\hline
\end{tabularx}
\begin{flushleft}
\vspace{2pt}\footnotesize $\bar{\alpha}$: mean calibration;\quad $\hat{\alpha}$: minmax calibration.\quad $^\dagger$BF16/FP64: BF16 model weights, FP64 sampling (reference software configuration).
\end{flushleft}
\end{table}

Under our best DART-BAOS configuration, KV4 quantization exceeds the BF16/FP64 baseline: $+1.9$\,pp on GSM8K (0-shot) and $+1.2$\,pp on HumanEval in prefix-cache mode, and $+1.8$\,pp on GSM8K (1-shot) and $+1.2$\,pp on HumanEval in dual-cache mode. By contrast, rotation-based smoothing~\cite{quarot} yields inconsistent results: it improves 0-shot GSM8K in prefix-cache mode but degrades HumanEval, and provides little benefit under dual-cache, highlighting the sensitivity of these AR-verified methods to diffusion-specific KV activation patterns.
On the sampling side, MXFP8 precision is broadly viable: under dual-cache, it surpasses the FP64 baseline on HumanEval by $+4.3$\,pp, likely because reduced-precision token selection introduces stochastic variation that occasionally benefits generation diversity. 
Overall, our full-stack 4-bit KV, 4-bit weight, 8-bit activation, and 16-bit sampling quantization remains competitive with the BF16/FP64 baseline across all evaluated benchmarks.

\subsection{Hardware Performance}
\label{subsec:e2e}

\begin{figure}[t]
    \centering
    \includegraphics[width=1.0\linewidth]{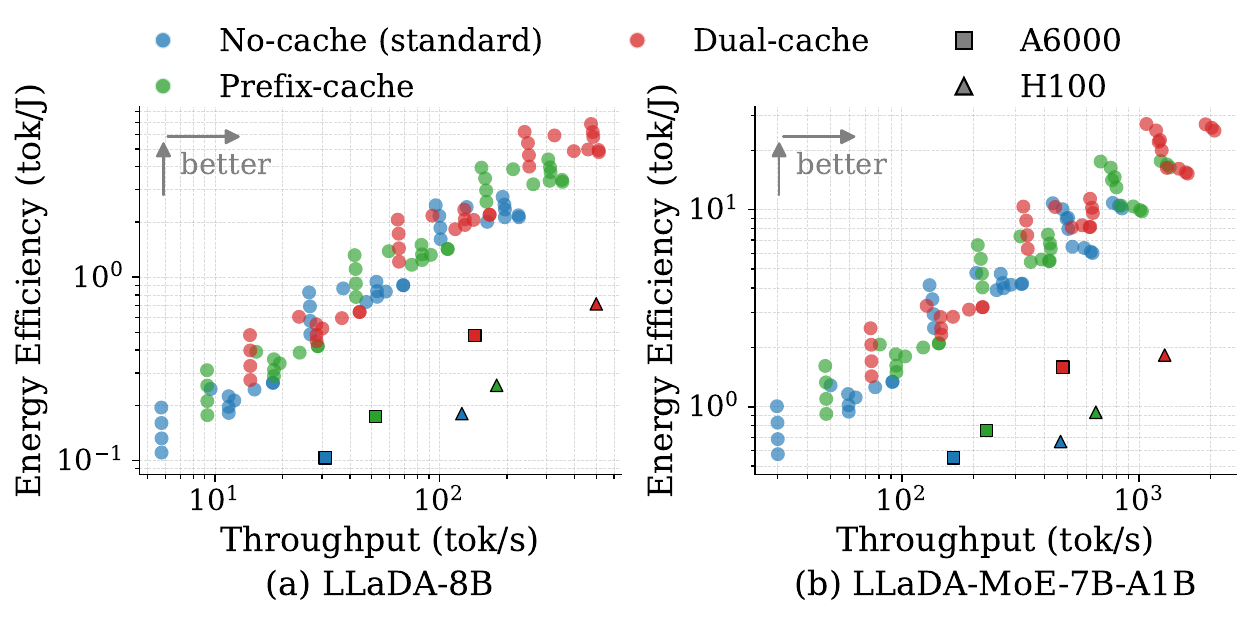}
    \caption{DART design-space sweep on dense and MoE diffusion models vs.\ GPU baselines.}
    \label{fg:dse}
\end{figure}

Fig.~\ref{fg:dse} sweeps DART hardware configurations ($VLEN \in$ \{256, 512, 1024, 2048\}, $MLEN \in$ \{256, 512, 1024\}, $BLEN \in$ \{4, 16, 64\}) across three inference paradigms and plots them against NVIDIA A6000 and H100 GPU. The workload is steps=$16$, block\_length=$64$, gen\_len=$256$, $B=16$ (IFEval prompts); DART configuration achieves higher tok/J than either GPU on the same throughput vertical line. Table~\ref{tab:e2e_main} reports detailed results at the $BLEN=64$, $VLEN=2048$, $MLEN=512$ operating point; TPS speedup and tok/J gain are relative to A6000 ($\times$1) within each model/cache block.

DART achieves $\times$4.91 TPS speedup over A6000 on LLaDA-8B (Prefix) and $\times$5.90 on LLaDA-8B (None), outperforming H100 across almost all configurations. Under the None cache paradigm, DART delivers $\times$5.90 and $\times$5.83 TPS speedup over A6000 for LLaDA-8B and LLaDA-MoE respectively. The energy efficiency advantage is substantial across all paradigms: DART delivers $\times$22.7--$\times$22.9 tok/J over A6000 on LLaDA-8B and $\times$18.4--$\times$19.7 on LLaDA-MoE. As for the area efficiency, DART achieves a compute area of 0.237 \(\mathrm{mm}^2\) and 27.83 TOPs/mm$^2$ calibrated at 4096 PEs.

\section{Related Work}
\label{se:RelatedWork}

\textbf{NPU accelerators for LLM inference.}
A decade of NPU and systolic-array research has been shaped by the AR inference paradigm. Designs such as FIGNA~\cite{figna}, OLIVE~\cite{olive}, and TENDER~\cite{tender} optimize for GEMM throughput under token-sequential decoding with append-only KV caching; MicroScopiQ~\cite{microscopiq} and PLENA~\cite{plena} co-design weight quantization with MX data formats for systolic hardware. Analytical and cycle-accurate simulation frameworks such as Timeloop~\cite{timeloop}, MAESTRO~\cite{maestro}, SCALE-Sim~\cite{scale-sim}, and ONNXim~\cite{onnxim} characterize GEMM dataflows but carry no model of non-GEMM sampling operations or diffusion-specific KV cache dynamics. None of these hardware designs address the non-GEMM sampling stage, block-wise KV refresh, or multiple-step refinement denoising in bi-directional attention that characterizes dLLM inference, and no dedicated NPU for dLLM inference exists.

\textbf{Diffusion language models and inference.} Masked diffusion LLMs such as LLaDA~\cite{llada}, DREAM~\cite{ye2025dream}, and MDLM~\cite{lou2023discrete} reformulate generation as iterative denoising of a fully masked sequence, achieving competitive quality with AR models on reasoning and instruction-following benchmarks. Block-wise decoding variants~\cite{arriola2025block,wu2025fast,wu2025fast-v2} accelerate inference by processing fixed-size token blocks with selective KV reuse, introducing the warm-step recomputation that BAOS exploits for calibration. On the software side, dInfer~\cite{dinfer} builds a vLLM-based serving framework tailored to dLLM workloads, optimizing GPU throughput through kernel fusion and scheduling.

\textbf{LLM quantization.} Post-training quantization (PTQ) for AR LLMs has been extensively studied across weight, activation, and KV cache quantization. For weight quantization, GPTQ~\cite{gptq}, AWQ~\cite{awq}, and OmniQuant~\cite{shao2023omniquant} represent widely adopted approaches. For activation and KV cache quantization, SmoothQuant~\cite{smoothquant} and QuaRot~\cite{quarot} address channel-wise outliers through per-channel scaling and Hadamard rotation, respectively, while P3-LLM~\cite{p3llm} computes smoothing factors during the AR prefill phase and reuses them throughout decoding. Recent work has also explored quantization and hardware co-design: MicroScopiQ~\cite{microscopiq} and PLENA~\cite{plena} co-design PTQ with microscaling (MX) data formats for systolic-array accelerators. These methods were designed under the static activation distributions characteristic of AR inference.

\textbf{Quantization for diffusion language models.} As pretrained diffusion language models gain traction, recent work has begun adapting PTQ techniques from the AR literature to this setting. \cite{lin2025quantization} is among the earliest to analysis the direct applciation of state-of-the-art PTQ to dLLMs, evaluating primarily on LLaDA and DREAM. DLLMQuant~\cite{xu2025dllmquant} proposes a fine-grained quantization approach that adapts calibration and error compensation to dynamic masking and iterative decoding, while Quant-dLLM~\cite{zhang2025quant} introduces a sensitivity-based precision allocation scheme that adaptively assigns bit widths across channel groups. However, these efforts focus on standard dLLM inference and do not address KV cache quantization under blocked diffusion decoding. In Section~\ref{se:Quantization}, we adapt several AR PTQ methods~\cite{gptq,quarot,p3llm,plena} to this setting and show that step-wise KV distribution shifts cause significant degradation. To our knowledge, DART is the first to apply microscaling data formats to dLLM quantization and the first to jointly address weight, activation, and KV cache quantization under blocked diffusion inference within a hardware co-designed framework.

\begin{table}[t]
\centering
\caption{End-to-end inference: A6000$^\dagger$, H100$^\dagger$, and DART$^\ddagger$. TPS speedup and tok/J gain relative to A6000 ($\times$1) per model/cache block.}
\label{tab:e2e_main}
\renewcommand{\arraystretch}{1.0}
\footnotesize
\setlength{\tabcolsep}{2pt}
\begin{tabular}{cccrrccc}
\toprule
\textbf{Model} & \textbf{Cache} & \textbf{Config} & \shortstack{\textbf{Total}\textbf{(s)}} & \textbf{TPS} & \shortstack{\textbf{Samp.} \textbf{(s,\,\%)}} & \shortstack{\textbf{TPS}} & \shortstack{\textbf{tok/J}\textbf{(×A6000)}} \\
\midrule
\multirow{9}{*}{\rotatebox[origin=c]{90}{LLaDA-8B}}
  & \multirow{3}{*}{None}
    & GPU (A6000)              & 190.97 & 31   & 3.93\,{\scriptsize(2.0\%)}  & $\times$1.00  & $\times$1.00  \\
  & & GPU (H100)               & 40.03  & 126  & 0.45\,{\scriptsize(1.1\%)}  & $\times$4.06  & $\times$1.74  \\
\cmidrule(l){3-8}
  & & DART                     & 22.32  & 183  & 0.15\,{\scriptsize(0.7\%)}  & $\times$5.90  & $\times$22.7  \\
\cmidrule(l){2-8}
  & \multirow{3}{*}{Prefix}
    & GPU (A6000)              & 90.24  & 52   & 0.72\,{\scriptsize(0.7\%)}  & $\times$1.00  & $\times$1.00  \\
  & & GPU (H100)               & 31.43  & 180  & 0.42\,{\scriptsize(1.3\%)}  & $\times$3.46  & $\times$1.48  \\
\cmidrule(l){3-8}
  & & DART                     & 16.06  & 255  & 0.10\,{\scriptsize(0.6\%)}  & $\times$4.91  & $\times$22.9  \\
\cmidrule(l){2-8}
  & \multirow{3}{*}{Dual}
    & GPU (A6000)              & 37.95  & 144  & 0.60\,{\scriptsize(1.5\%)}  & $\times$1.00  & $\times$1.00  \\
  & & GPU (H100)               & 9.94   & 500  & 0.33\,{\scriptsize(3.4\%)}  & $\times$3.47  & $\times$1.49  \\
\cmidrule(l){3-8}
  & & DART                     & 10.77  & 380  & 0.06\,{\scriptsize(0.6\%)}  & $\times$2.64  & $\times$12.4  \\
\midrule
\multirow{9}{*}{\rotatebox[origin=c]{90}{LLaDA-MoE-7B}}
  & \multirow{3}{*}{None}
    & GPU (A6000)              & 39.76  & 165  & 0.79\,{\scriptsize(2.0\%)}  & $\times$1.00  & $\times$1.00  \\
  & & GPU (H100)               & 13.18  & 466  & 0.42\,{\scriptsize(3.2\%)}  & $\times$2.82  & $\times$1.21  \\
\cmidrule(l){3-8}
  & & DART                     & 4.26   & 962  & 0.19\,{\scriptsize(4.6\%)}  & $\times$5.83  & $\times$18.4  \\
\cmidrule(l){2-8}
  & \multirow{3}{*}{Prefix}
    & GPU (A6000)              & 26.30  & 227  & 0.77\,{\scriptsize(2.9\%)}  & $\times$1.00  & $\times$1.00  \\
  & & GPU (H100)               & 9.25   & 656  & 0.37\,{\scriptsize(4.0\%)}  & $\times$2.89  & $\times$1.24  \\
\cmidrule(l){3-8}
  & & DART                     & 4.39   & 932  & 0.12\,{\scriptsize(2.7\%)}  & $\times$4.11  & $\times$19.7  \\
\cmidrule(l){2-8}
  & \multirow{3}{*}{Dual}
    & GPU (A6000)              & 12.74  & 476  & 0.63\,{\scriptsize(4.7\%)}  & $\times$1.00  & $\times$1.00  \\
  & & GPU (H100)               & 4.77   & 1279 & 0.29\,{\scriptsize(6.1\%)}  & $\times$2.69  & $\times$1.15  \\
\cmidrule(l){3-8}
  & & DART                     & 2.81   & 1456 & 0.08\,{\scriptsize(2.7\%)}  & $\times$3.06  & $\times$14.6  \\
\bottomrule
\end{tabular}
\begin{flushleft}
\vspace{2pt}\footnotesize
$^\dagger$\,GPU rows measured via the dInfer framework~\cite{dinfer} (vLLM backend), BF16 model weights and BF16 sampling precision.\\
$^\ddagger$\,DART rows use the full-stack quantization configuration of Table~\ref{tab:v1_quant}: MXINT4 KV cache and weights, MXINT8 activations, BF16 sampling.
\end{flushleft}
\end{table}
\section{Conclusion}
\label{se:Conclusion}

We presented DART, the first configurable NPU platform for dLLM inference, providing a full-stack design that spans the transformer forward pass, diffusion sampling, block-wise KV caching, and hardware-friendly quantization. The platform was validated through a tri-path simulation framework cross-validated against AMD HBM2e physical measurements and Verilator RTL, and synthesized at 7\,nm. Evaluation against A6000 and H100 GPUs confirms that DART ($\mathit{VLEN}{=}2048$) achieves up to ${\times}4.91$ TPS speedup and ${\times}23.3$ energy efficiency over A6000 while surpassing H100 across all configurations. These results demonstrate that dedicated hardware support for dLLM-specific workload patterns, namely bidirectional full-sequence passes, non-monotonic KV refresh, and reduction-heavy sampling, translates directly into measurable efficiency gains over general-purpose GPU execution.

Taken together, DART establishes both an efficient and scalable diffusion language model-specific NPU inference stack for the community.

\clearpage

\bibliographystyle{ACM-Reference-Format}
\bibliography{main}

\end{document}